\documentclass[11pt]{article}
\usepackage{moriond,epsfig}

\def\Journal#1#2#3#4{{#1} {\bf #2}, #3 (#4)}

\def\PLB{{\em Phys. Lett.}  B}
\def\PRL{\em Phys. Rev. Lett.}
\def\PRD{{\em Phys. Rev.} D}

\def\be{\begin{equation}}
\def\ee{\end{equation}}
\def\bea{\begin{eqnarray}}
\def\eea{\end{eqnarray}}

\begin{document}
\vspace*{4cm}
\title{HIGH ENERGY PHYSICS AND QUINTESSENCE\footnote{CERN-TH/2000-078}}

\author{P. BRAX${}^{*}$ and J. MARTIN${}^{\dagger }$}

\address{${}^{*}$ CERN, Theory division, Gen\`eve, Switzerland,\\
and Service de Physique Th\'eorique, CEA-Saclay, F-91191 Gif/Yvette Cedex, \\
${}^{\dagger }$DARC, Observatoire de Paris, 
UMR 8629 CNRS, 92195 Meudon Cedex, France.}

\maketitle
\abstracts{It is shown that any realistic model of quintessence should be 
based on Supergravity (SUGRA) since the value of the quintessence field on 
the attractor is approximately the Planck mass. Under very general assumptions, 
the typical shape of a SUGRA tracking potential is derived. Cosmological 
implications are investigated. In particular, it is demonstrated that, generically, the 
equation of state parameter is driven to a value close to $-1$ in 
agreement with recent observations.}

If confirmed, the discovery that our Universe is presently 
undergoing a phase of accelerated expansion \cite{BOPS} has clearly important 
implications for Cosmology and High Energy Physics. In particular, 
a crucial question is to identify the physical nature of the 
dark matter component responsible for this acceleration which 
would represent $70 \%$ of the matter content of the Universe. A natural 
candidate is obviously the cosmological constant. The cosmological 
constant energy density includes in fact contributions from two 
origins. The first contribution is due to the bared 
cosmological, $\Lambda _0$, which appears in the Einstein equations 
as a ``free parameter'' and the second one is due to the zero-point 
fluctuations of the quantum fields. This gives rise to energy densities 
respectively equal to $\Lambda _0c^4/(8\pi G)$ and 
$(\hbar c/2)\int {\rm d}^3 {\bf k}k/(2\pi )^3\approx \hbar ck_{\rm max}^4
/(16\pi ^2)$. In the last expression, $k_{\rm max}$ is a cut-off which 
can naturally be taken as the Planck wavenumber. As a consequence, the 
second contribution becomes huge. Since the observed value is tiny ($0.7$ 
times the critical energy density $\rho _{\rm c}$), one needs to fine-tune 
very accurately the value of the bared constant $\Lambda _0$ such that 
the previous simple theoretical predictions be compatible with observational 
data. Let us emphasize that the discovery that the Universe is accelerating renders 
this problem even worse than before. Indeed, it is probably easier to 
explain why there is an exact cancellation between the two contributions 
such that $\rho _{\Lambda }=0$ (due, for example, to some so far unknown 
symmetry in quantum gravity) than finding a reason for a non-vanishing 
tiny value $\rho _{\Lambda }=0.7\times \rho _{\rm c}\neq 0$.
\par
This problem motivates the search for alternative explanations. Among 
them is the quintessence hypothesis. It consists in assuming that 
the dark matter component is an homogeneous scalar field $Q(t)$, 
named quintessence, such that $\Omega _{\rm Q}=0.7$ (and therefore 
$\Omega _{\Lambda }=0$). However, this is not sufficient to be an 
interesting proposal. In addition, the shape of the potential $V(Q)$ must 
possess some typical features. Roughly speaking, the parameter $\Gamma $ defined 
by the expression $\Gamma \equiv ({\rm d}^2V(Q)/{\rm d}Q^2)V(Q)/({\rm d}V(Q)/{\rm d}Q)^2$ 
must be greater than one and almost constant \cite{SWZ}. The prototype of such 
a potential is given by $V(Q)=\Lambda ^{4+\alpha }/Q^{\alpha }$ \cite{RP,CDS} where 
$\Lambda $ and $\alpha >0$ are two free parameters. 
\par
Let us now examine briefly what are the main properties of this ``tracker'' 
quintessence field. In order to have $\Omega _{\rm Q}=0.7$, we must 
tune the value of $\Lambda $. For example, if $\alpha =6$, one has 
$\Lambda \approx 10^6\mbox{GeV}$, a scale compatible with the typical 
scales of High Energy Physics. In this sense, the fine-tuning problem is 
less important than for the cosmological constant since the free parameter 
of the potential can now take a ``natural'' value. This would not have been 
the case for the potential $V(Q)=(1/2)m^2Q^2$ where the only free parameter 
must be chosen such that $m\approx 10^{-33}\mbox{eV}$, certainly 
an ``unnatural'' value. For the 
tracking potential the situation is different since the mass, defined as the second 
derivative of the potential, has also this small value; but, as we have seen, this is not 
the result of an artificial choice. This is rather the consequence of the shape 
of the potential and of the reasonable value of the free parameter $\Lambda $. Therefore, 
even if the problem has not been completely overcome, something has been gained. 
\par
Having fixed the value of $\Lambda $ and $\alpha $, we now turn to the study of the 
dependence of the final result on the initial conditions. This is the most interesting 
property of the tracking potential. It can be shown that, due to the presence of 
an attractor, the final result is completely independent of the initial conditions. The 
initial value of the energy density can vary in a range of $100 $ orders 
of magnitude: $10^{61}\mbox{GeV}^4< \rho _{\rm Q}< 10^{-37}\mbox{GeV}^4$. Another 
nice property of such models is that it turns out that the equation 
of state parameter, $\omega _{\rm Q}\equiv p_{\rm Q}/\rho _{\rm Q}$, is naturally 
such that $-1<\omega _{\rm Q}< 0$. However, it is not possible to 
obtain a value less than $\omega _{\rm Q}=-0.7$ which could be a problem since 
observations seem to suggest a value close to $-1$ \cite{Ef}. 
\par
\begin{figure}
\center
\psfig{figure=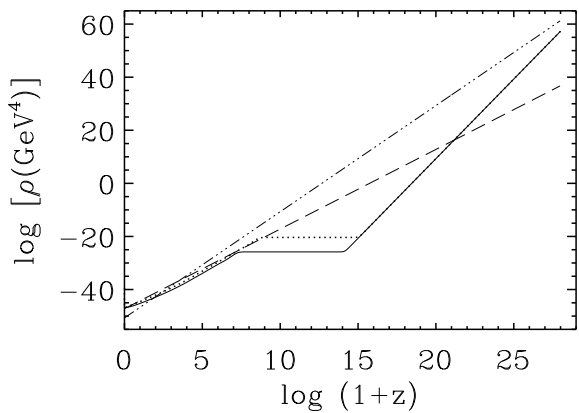,height=2.5in}
\caption{Evolution of the different energy densities. The dashed-dotted line represents the 
the energy density of radiation whereas the dashed line represents the energy 
density of matter. The solid line is the energy density of quintessence in the SUGRA 
model with $\alpha =11$. The dotted line is the energy density of quintessence 
for the potential $V(Q)=\Lambda ^{4+\alpha }Q^{-\alpha }$ with the same 
$\alpha $. The initial conditions are such that equipartition, i.e. $\Omega _{\rm Qi}
\approx 10^{-4}$ is realized just after inflation.
\label{energydensity}}
\end{figure}

One of the most interesting 
challenge is to derive the tracking properties from generic High Energy 
Physics considerations. We now show briefly that this is indeed possible and 
that this leads to a different shape for the potential for which all the previous 
nice properties are preserved. In addition  some of the problems 
of the previous approach are overcome. Let us consider several possible options for the 
particle physics origin of the quintessence field. For that it is relevant to express the mass
of the quintessence field in terms of the Hubble constant
\begin{equation}
\label{attractor}
\frac{{\rm d}^2 V(Q)}{{\rm d}^2 Q}=\frac{9}{2}\frac{1+\alpha}{\alpha}(1-\omega_{\rm Q}^2)H^2.
\end{equation}
This has two immediate consequences. First of all, as already mentioned above, the mass of 
the quintessence field is extremely light compared to the particle physics scales.
This implies that the quintessence field could mediate an extremely long range
fifth force. This violates experimental results entailling that the quintessence
field must be decoupled from ordinary matter. On the other hand, the value
of $Q$ now (at vanishing redshift) is of order $m_{\rm Pl}$ as can be seen on the previous 
formula. This strongly suggests that High 
Energy Physics must be taken into account in trying to find the origin of the quintessence 
field. We will only consider models beyond the standard model
of particle physics which incorporate supersymmetry. Moreover we shall explicitly consider 
the supergravity effects as they cannot be neglected for $Q\approx m_{\rm Pl}$. Within this 
scheme several possibilities can be envisaged. The quintessence field could 
spring from the meson field after the gauge breaking of supersymmetric gauge 
theory \cite{Bine}, it could also be the dilaton or one of the compactication moduli.
It is easy to see that the supergravity potential of the meson field leads to 
nonsensical negative energy densities. The potential for the dilaton $S$ is of the
type $\exp (\exp -S)$ which is far too flat. The potential for the moduli $T$
is of exponential type $\exp (-\lambda T)$ which leads to a small value
of $\Omega_{\rm Q}\le 0.15$ \cite{FJ}. We are thus lead to postulate the existence of 
a quintessence field whose physical origin is not clear from the string point
of view. Nevertheless, one can adopt an effective theory point of view
and look for a general class of supegravity models leading to a quintessence
behaviour \cite{BM1,BM2}. For that it is only necessary to specify two functions,
the superpotential $W$ which is a holomorphic function of the fields and
the K\"ahler potential governing the kinetic terms. The Lagrangian depends on 
the combination $G= \kappa K+\ln(\kappa^3  \vert W\vert ^2)$ where
$\kappa\equiv 8\pi/m_{\rm Pl}^2$. The scalar potential is given by $V=\kappa^{-2}e^G
(G^iG_i-3)+V_{\rm D}$ where $V_{\rm D}$ is a positive contribution from the gauge 
sector of the models. From this one can analyse the typical behaviour of the 
scalar potential in a supergravity quintessence model. First of all
for a general polynomial $K$ the exponential term $e^G$ will be negligible
throughout the cosmic evolution of $Q$ at least until $Q\approx m_{\rm Pl}$. 
This implies that one looks for models where the $G^iG_i$ term is responsible
for the tracking potential. It is only at low red shift that the $e^G$ 
supergravity correction starts contributing by slowing the evolution of $Q$
hence pushing the value of the equation of state towards $-1$. 
In Refs. \cite{BM1,BM2} string inspired K\"ahler and superpotentials were proposed along these
lines. They lead to the SUGRA tracking potential of the form
\begin{equation}
\label{sugrapot}
V(Q)=\frac{\Lambda^{4+\alpha}}{Q^{\alpha}}e^{\frac{\kappa}{2}Q^2}.
\end{equation}
The cosmological implications of these potentials follow. First of all, the 
value of $\Lambda $ necessary to obtain $\Omega _{\rm Q}\approx 0.7$ is still the same, i.e. 
comparable to the scales of particles physics. Secondly, the mass of this potential 
is now given by $m\approx \sqrt{\rho _{\rm c}}/m_{\rm Pl}e^{2\pi }$. It is therefore 
still very small even if we have gained a factor $e^{2\pi }$. Thirdly, we see that the 
coincidence problem is still solved since, during almost all the cosmic evolution, the 
exponential factor plays no role. Hence, the attractor is still present. This 
behaviour is illustrated in  Fig. (\ref{energydensity}) where the evolution of the 
energy density of the quintessence field is displayed.

\begin{figure}
\center
\psfig{figure=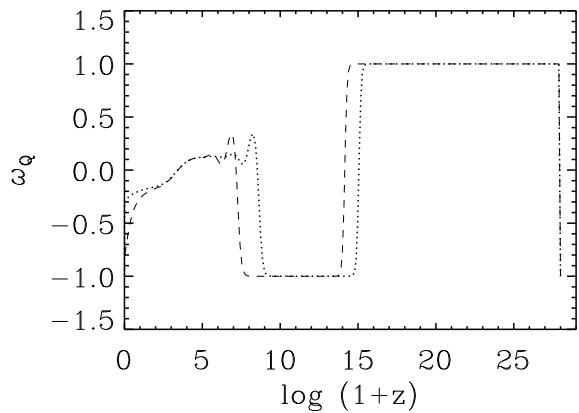,height=2.5in}
\caption{The dotted line represents the evolution of $\omega _{\rm Q}$ 
for the potential given by $V(Q)=\Lambda ^{4+\alpha }Q^{-\alpha }$ with 
$\alpha =11$ whereas the dashed line represents the evolution of $\omega _{\rm Q}$ for the 
SUGRA tracking potential with the same value of $\alpha $.
\label{eqstate}}
\end{figure}

As already mentioned above, the most interesting cosmological implications 
concern the equation of state parameter. Its evolution is displayed in 
Fig. (\ref{eqstate}). It is clear that the equation of state parameter is driven 
towards $-1$ at very small redshifts due to the presence of the exponential 
factor. Numerical integration shows that, for the SUGRA tracking potential, its 
precise value is given by
\begin{equation}
\label{omegaQ}
\omega _{\rm Q}\approx -0.82.
\end{equation}
It should be emphasized that this value is independent of the free parameter 
$\alpha $.

%This property is illustrated in Fig. (\ref{aa})

%\begin{figure}
%\center
%\psfig{figure=oo.ps,height=2.5in}
%\caption{$\omega _{\rm Q}-\alpha $ relation for the SUGRA tracking 
%potential
%\label{aa}}
%\end{figure}

In conclusion, we have shown in this article that a realistic model 
of quintessence must be based on SUGRA since the value of the field 
is $\approx m_{\rm Pl}$ on the attractor. There are of course questions 
which remain to be addressed, in particular the issue of SUSY 
breaking \cite{BM3}.

\end{document}